\documentclass[preprint,prd,aps,nofootinbib]{revtex4}
\usepackage{amsmath,graphics}
\begin{document}
\begin{center}
\bibliographystyle{article}
{\Large \textsc{Quantum cosmology from three different perspectives}}
\end{center}
\vspace{0.4cm}

\author{Giampiero Esposito$^{1,2}$ \thanks{
Electronic address: giampiero.esposito@na.infn.it}}

\affiliation{
${\ }^{1}$Istituto Nazionale di Fisica Nucleare, Sezione di Napoli,\\ 
Complesso Universitario di Monte S.
Angelo, Via Cintia, Edificio N', 80126 Naples, Italy\\
${\ }^{2}$Dipartimento di Scienze Fisiche, Complesso Universitario di
Monte S. Angelo,\\ 
Via Cintia, Edificio N', 80126 Naples, Italy}

\begin{abstract}
Our review is devoted to three promising research lines in quantum cosmology
and the physics of the early universe. The nonperturbative renormalization 
programme is making encouraging progress that we here assess from the point
of view of cosmological applications: Lagrangian and Hamiltonian form of
pure gravity with variable $G$ and $\Lambda$; power-law inflation for pure
gravity; an accelerating universe without dark energy. In perturbative 
quantum cosmology, on the other hand, diffeomorphism-invariant boundary
conditions lead naturally to a singularity-free one-loop wave function of
the Universe. Last, but not least, in the braneworld picture one discovers the
novel concept of cosmological wave function of the bulk space-time. Its impact
on quantum cosmology and singularity avoidance is still, to a large extent,
unexplored.
\end{abstract}
\maketitle
\bigskip
\vspace{2cm}

\section{Foreword}
In this brief review we focus on three (among the many) peculiar aspects
of modern quantum cosmology, with the hope of leading quickly the reader
towards open research problems. No completeness can be achieved in such 
a short presentation, and we therefore apologize in advance with the 
colleagues who might not find a proper acknowledgment of their work.

\section{Quantum cosmology via functional integrals}

The familiar formulation of quantum cosmology via functional integrals
relies upon the pioneering work of Misner \cite{Misn57}, Hartle and
Hawking \cite{Hart83}. The main idea of the functional-integral approach
is to build in-out amplitudes following Feynman: the amplitude to go from
a metric $g_{1}$ and a (matter) field configuration $\phi_{1}$ on a 
spacelike surface $S_{1}$ to a metric $g_{2}$ and a (matter) field
configuration $\phi_{2}$ on a spacelike surface $S_{2}$ is (formally) 
expressed as the functional integral of the exponential of $i$ times 
the action, supplemented by gauge-fixing and ghost 
terms \cite{Dewi03}, taken over all metrics and (matter) fields matching 
the given boundary data on $S_{1}$ and $S_{2}$. In order to obtain a
well-defined prescription, in-out amplitudes are sometimes written first 
as Euclidean functional integrals, but severe technical 
problems occur: integration 
measure over all four-geometries with their topologies 
and unboundedness from below of the
Euclidean action among the many \cite{Hawk79}.

\section{Hartle--Hawking quantum state}

In a cosmological setting, one therefore arrives at the 
Hartle--Hawking quantum state \cite{Hart83}. According to these authors,
the quantum state of the Universe \cite{Hawk84} 
can be expressed by an Euclidean functional
integral over compact four-geometries matching the boundary data on the
surface $S_{2}$, while the three-surface $S_{1}$ shrinks to a point
(hence the name ``no boundary proposal''). One can therefore derive,
in principle, all we know about cosmology from a choice of boundary
conditions \cite{Hawk82}, including formation of 
structure \cite{Hall85}, coupling 
to matter fields \cite{Deat87}, inflationary solutions \cite{Espo88}, 
supersymmetric models \cite{Deat96, Moni96}.

\section{Renormalization-group approach}

Recent progress relies instead on a completely different approach: one
builds a scale-dependent effective action $\Gamma(k)$ for quantum Einstein
gravity, which is ruled by the renormalization-group 
(hereafter RG) equation. If
$\Gamma(k)$ equals the classical Einstein--Hilbert action at the ultraviolet
cut-off scale $\kappa$, one uses the RG equation to evaluate $\Gamma(k)$
$\forall k < \kappa$, and then sends $k \rightarrow 0$ and 
$\kappa \rightarrow \infty$. The continuum limit as $\kappa \rightarrow
\infty$ should exist after renormalizing finitely many parameters in the
action, and is taken at a non-Gaussian fixed point of the RG-flow
\cite{Bona02}. Over the years, strong evidence has been obtained in favour
of the new ultraviolet fixed point, regardless of the trunction used
\cite{Laus05}. The plot in \cite{Laus05} shows part of theory space of the
Einstein--Hilbert truncation with its RG flow. The arrows therein point in 
the direction of decreasing values of $k$ \cite{Espo06}. The flow is
dominated by a non-Gaussian fixed point in the first quadrant and a trivial
one at the origin \cite{Espo06}. 

\section{Cosmological applications}

After investigating the RG-improved equations for self-interacting scalar
fields coupled to gravity in a FLRW Universe \cite{Bona03}, we have improved
the action principle itself, building the Lagrangian and Hamiltonian 
formalism with variable $G$ and $\Lambda$ 
{\it treated as dynamical variables} \cite{Bona04}. The latter point is
substantially innovative, since all other investigations in the literature
treated $G$ and $\Lambda$ as external parameters at the very best, but not
as dynamical variables with an Euler--Lagrange equation for $G$.

\section{Power-law inflation for pure gravity}

Unlike models where only the Einstein equations are RG-improved, our 
framework allows for a non-trivial dynamics of the scale factor even in the
absence of coupling to a matter field. Indeed, if in the pure-gravity 
case we look for power-law solutions of the Euler--Lagrange equations 
of the type \cite{Bona04, Bona05}
\begin{equation}
a(t)=At^{\alpha}, \; G(t)=g_{\star}{t^{2}\over \xi^{2}}, \;
\Lambda(t)=\lambda_{\star}{\xi^{2}\over t^{2}},
\label{(1)}
\end{equation}
we find for example, in a spatially flat FLRW Universe, that $A$ is
undetermined, while
\begin{equation}
\alpha={1\over 6}\left(3 \pm \sqrt{9+12 \xi^{2}\lambda_{\star}}\right).
\label{(2)}
\end{equation}
Our modified Lagrangian \cite{Bona04} allows therefore for power-law
inflation in pure-gravity models, unlike all previous models in the
literature \cite{Barv93, Laus05, Baue06}.

\section{Infrared fixed point}

The derivation of an infrared fixed point is not on a footing as firm
as the evidence in favour of an ultraviolet fixed point \cite{Laus05}.
Nevertheless, on assuming its existence, we have linearized the RG-flow
and, after evaluating the critical exponents, we have found how the
infrared fixed point would be approached \cite{Bona06}. We have also
obtained a smooth transition between FLRW cosmology and the observed
accelerated expansion of the universe \cite{Bona06}.

\section{Perturbative quantum cosmology}

Perturbative quantum cosmology studies instead the first quantum 
corrections to the underlying classical dynamics. In particular, one-loop
effects can be evaluated after imposing gauge-invariant boundary conditions,
according to the recipe for imposing gauge-invariant boundary conditions
in quantum field theory \cite{Avra99}. On denoting by $\pi_{\; j}^{i}$
a projector acting on the gauge fields $\varphi^{j}$, by 
$P^{\alpha}(\varphi)$ and $\psi_{\beta}$ the gauge-fixing functionals 
and ghost fields, respectively, such boundary conditions read as
\begin{equation}
\left[\pi_{\; j}^{i}\; \varphi^{j}\right]_{\partial M}=0,
\label{(3)}
\end{equation}
\begin{equation}
\left[P^{\alpha}(\varphi)\right]_{\partial M}=0,
\label{(4)}
\end{equation}
\begin{equation}
[\psi_{\beta}]_{\partial M}=0.
\label{(5)}
\end{equation}

\section{Singularity avoidance at one loop?}

For pure gravity, one-loop quantum cosmology in the limit of small
three-geometry \cite{Schl85} describes a vanishing probability of
reaching the singularity at the origin (of the Euclidean four-ball) 
only with diffeomorphism-invariant boundary conditions
\cite{Espo05a, Espo05b}, which are a particular case of the previous
scheme. All other sets of boundary conditions lead instead to a
divergent one-loop wave function \cite{Schl85, Espo97, Espo05b}.

\section{Peculiar property of the four-ball?}

We stress that we do not require a vanishing one-loop wave function.
We rather find it, on the Euclidean four-ball, as a consequence of
diffeomorphism-invariant boundary conditions. Peculiar cancellations 
occur on the Euclidean four-ball, and the spectral (also called 
generalized) $\zeta$-function remains regular at the origin
\cite{Espo05a, Espo05b}, despite the lack of strong ellipticity
of the boundary-value problem \cite{Avra99}.

\section{Towards brane-world quantum cosmology}

In the braneworld picture, branes are timelike surfaces with metric
$g_{\alpha \beta}$ embedded into bulk space-time with metric
$G_{AB}$. The action functional can be taken to be the sum of a
four-dimensional (brane) and five-dimensional (bulk) contribution, 
i.e. \cite{Barv05}
\begin{equation}
S=S_{4}[g_{\alpha \beta}(x)]+S_{5}[G_{AB}(X)].
\label{(6)}
\end{equation}
In general, there exist vector fields $R_{B},R_{\nu}$ on the space of
histories such that 
\begin{equation}
R_{B}S_{5}=0, \; R_{\nu}S_{4}=0,
\label{(7)}
\end{equation}
with Lie brackets given by
\begin{equation}
[R_{B},R_{D}]=C_{\; BD}^{A} \; R_{A}, \;
[R_{\mu},R_{\nu}]=C_{\; \mu \nu}^{\lambda} \; R_{\lambda}.
\label{(8)}
\end{equation}
The components of the vector fields $R_{B}$ and $R_{\nu}$ generate
five-dimensional and four-dimensional diffeomorphisms, respectively, 
while the bulk and brane ghost operators read 
\cite{Barv05} (with $F^{A}$ and
$\chi^{\mu}$ the bulk and brane gauge-fixing functionals, respectively)
\begin{equation}
Q_{\; B}^{A} \equiv R_{B}F^{A}=F_{\; ,a}^{A} \; R_{\; B}^{a},
\label{(9)}
\end{equation}
\begin{equation}
J_{\; \nu}^{\mu} \equiv R_{\nu}\chi^{\mu}=\chi_{\; ,i}^{\mu}
\; R_{\; \nu}^{i}.
\label{(10)}
\end{equation}
On denoting by $S_{A}$ and $T^{B}$ the bulk ghost fields, the 
cosmological wave function of the bulk space-time 
can be written as \cite{Barv05}
\begin{equation}
\psi_{\rm Bulk}=\int_{G_{AB}[\partial M]=g_{\alpha \beta}}
\mu(G_{AB},S,T)e^{{\rm i}{\widetilde S}_{5}},
\label{(11)}
\end{equation}
with $\mu(G_{AB},S,T)$ a suitable measure functional, while
\begin{equation}
{\widetilde S}_{5}=S_{5}[G_{AB}]+{1\over 2}F^{A}\omega_{AB}F^{B}
+S_{A}Q_{\; B}^{A} T^{B}.
\label{(12)}
\end{equation}

\section{Braneworld effective action}

The braneworld effective action $\Gamma$ can (in principle) be
obtained from the formula \cite{Barv05}
\begin{equation}
e^{{\rm i}\Gamma}=\int \nu(g_{\alpha \beta},\rho,\sigma)
e^{{\rm i}{\widetilde S}_{4}}\psi_{\rm Bulk},
\label{(13)}
\end{equation}
where $\nu(g_{\alpha \beta},\rho,\sigma)$ is a suitable measure
functional over brane metrics and brane ghost fields, 
while \cite{Barv05}
\begin{equation}
{\widetilde S}_{4}=S_{4}+{1\over 2}\chi^{\mu}C_{\mu \nu}\chi^{\nu}
+\rho_{\mu} J_{\; \nu}^{\mu}\sigma^{\nu}.
\label{(14)}
\end{equation}
Recent developments in this respect can be found in \cite{Barv06}, where
the authors lay the foundations for a sistematic application of the
background-field method to the braneworld picture.

\section{Selected open problems}

In our opinion, it is of crucial importance to work at 
least on the following unsettled issues:
\vskip 0.3cm
\noindent
(i) Can one prove in a rigorous way that an infrared fixed point occurs
in the nonperturbative approach?
\vskip 0.3cm
\noindent
(ii) Can the spectral cancellations found in \cite{Espo05a, Espo05b}
survive the choice of curved backgrounds with boundary?
\vskip 0.3cm
\noindent
(iii) Is braneworld quantum cosmology one-loop singularity free?

Hopefully, the years to come will shed some light on these open problems,
and we also hope that quantum cosmology will make a closer contact with
the rich world of observational cosmology.

\acknowledgments

The work of G. Esposito has been partially supported by PRIN
{\it SINTESI}. Collaboration with I. Avramidi, A. Bonanno, G. Fucci,
A.Yu. Kamenshchik, K. Kirsten, C. Rubano, P. Scudellaro has been
very helpful in the course of working on the topics this short review
is devoted to.

\end{document}